\documentclass[times,10pt,onecolumn]{article}

\usepackage{epsf}\epsfverbosetrue
\usepackage{graphics,epsfig,color}
\usepackage{graphicx,epsfig} 
\usepackage{multirow}
\usepackage{alltt}
\usepackage{subfigure}
\usepackage{float}
\usepackage{url}
\usepackage{graphicx}
\usepackage{verbatim} 
\usepackage{algorithm}
\usepackage{algpseudocode}
\usepackage{footnote} 
\usepackage{latexsym} 
\usepackage{amsmath}
\usepackage{sidecap} 
\usepackage{color}

\begin{document}

\title{A Tutorial on Broadcasting Packets over Multiple-Channels in a Multi-Inferface Network Setting in NS-2\vspace{50pt}}

\author{Zeeshan Ali Khan and Mubashir Husain Rehmani \vspace{170pt} \thanks{Z. A. Khan is a PhD student at LEAT, CNRS, University of Nice Sophia Antipolis, Sophia Antipolis, France e-mail: zkhan@unice.fr; M. H. Rehmani is a PhD studnet at Lip6/Universti\'{e} Pierre et Marie Curie (UPMC) -- Paris-6, France, e-mail: mubashir.rehmani@lip6.fr. The author M. H. Rehmani would like to thanks A. C. Viana, who is with INRIA, France email: aline.viana@inria.fr} }

\date{Version 1 \vspace{60pt} \\ $5^{th}$ July 2010}

\maketitle
\pagebreak
\thispagestyle{plain}

\tableofcontents
\pagebreak

\begin{abstract}
With the proliferation of cheaper electronic devices, wireless communication over multiple-channels in a multi-interface network is now possible. For instace, wireless sensor nodes can now operate over multiple-channels. Moreover, cognitive radio sensor networks are also evolving, which also operates over multiple-channels. In the market, we can find antennas that can support the operation of multiple channels, for e.g. the cc2420 antenna that is used for communication between wireless sensor nodes consists of 16 programmable channels. The proper utilization of multiple-channels reduces the interference between the nodes and increase the network throughput. Recently, a Cognitive Radio Cognitive Network (CRCN) patch for NS-2 simulator has proposed to support multi-channel multi-interface capability in NS-2.

In this tutorial, we consider how to simulate a multi-channel multi-interface wireless network using the NS-2 simulator. This tutorial is trageted to the novice users who wants to understand the implementation of multi-channel multi-interface in NS-2. We take the Cognitive Radio Cognitive Network (CRCN) patch for NS-2 simulator and demonstrate broadcasting over multiple-channels in a multi-interface network setting. In our seeting, node braodcasts the Hello packets to its neighbors. Neighboring nodes receive the Hello packets if and only if they are tuned to the same channel. We demonstrate through example that the tuning of receivers can be done in two fashions. 
\end{abstract}


\section{Introduction}
\label{sec:introduction}

The Network Simulator (NS-2) \cite{IEEEhowto:ns} is a most widely used network simulator. This tutorial uses the implementation of WCETT~\cite{wcett} based multi-channel wireless mesh protocol in the Cognitive Radio Cognitive Network (CRCN)~\cite{crcn} patch developed for NS-2. The expected audience are students who want to understand the broadcasting mechanism in multi-channel multi-interface environment in NS-2. The version considered is NS-2.32 and 2.33, but it might be useful to other versions as well. Throughout the rest of this tutorial, the under considered files are wcett.cc, wcett.h, wcett\_logs.cc, wcett\_packet.h, wcett\_rqueue.cc, wcett\_rqueue.h, wcett\_rtable.cc, wcett\_rtable.h, which can be found in WCETT folder in the NS-2 base directory. The tcl file test4wcett.tcl can be found in the NS-2 based directory. Other relevant files are phy.cc and wireless-phy.cc.



\section{Methodology}
Our goal is to broadcast Hello packets over multiple interfaces. The methodology we adopted is to use Hello packets and broadcast them over multiple interfaces. We then demonstrated that a specific channel can be assigned to a particular interface. This assigned can be done either directly at the lower layer or can be communicated from the routing layer to the lower layer. We now describe the relevant section of codes. 


\subsection{Modifications in TCL file test4wcett.tcl}
We use the TCL file test4wcett.tcl present in the NS-2 folder. WCETT is used as a routing protocol, whose code is quite similar to AODV and having the additional capability of calculating WCETT metric and can be able to pass channel selection information to lower layers. IEEE 802.11 mac is selected as mac protocol. The total number of interfaces of each node is selected as 3 in the tcl script:

set val(ni) 3;

\subsection{How to Enable Hello Packets}
\label{sec:hello}

By default HELLO packets are disabled in the wcett protocol. To enable broadcasting of Hello packets, comment the following two lines present in wcett.cc\\ \#ifndef WCETT\_LINK\_LAYER\_DETECTION\\ \#endif LINK LAYER DETECTION 
and recompile NS-2 by using the following commands on the terminal:\\
make clean\\
make \\
sudo make install\\

\subsection{How to Send Hello Packets on a Specific Channel over Multiple Interfaces}

\vspace{-0.65cm}
\begin{figure}[h]
    \begin{center}
    {
        \includegraphics[width=12cm, height=7cm]{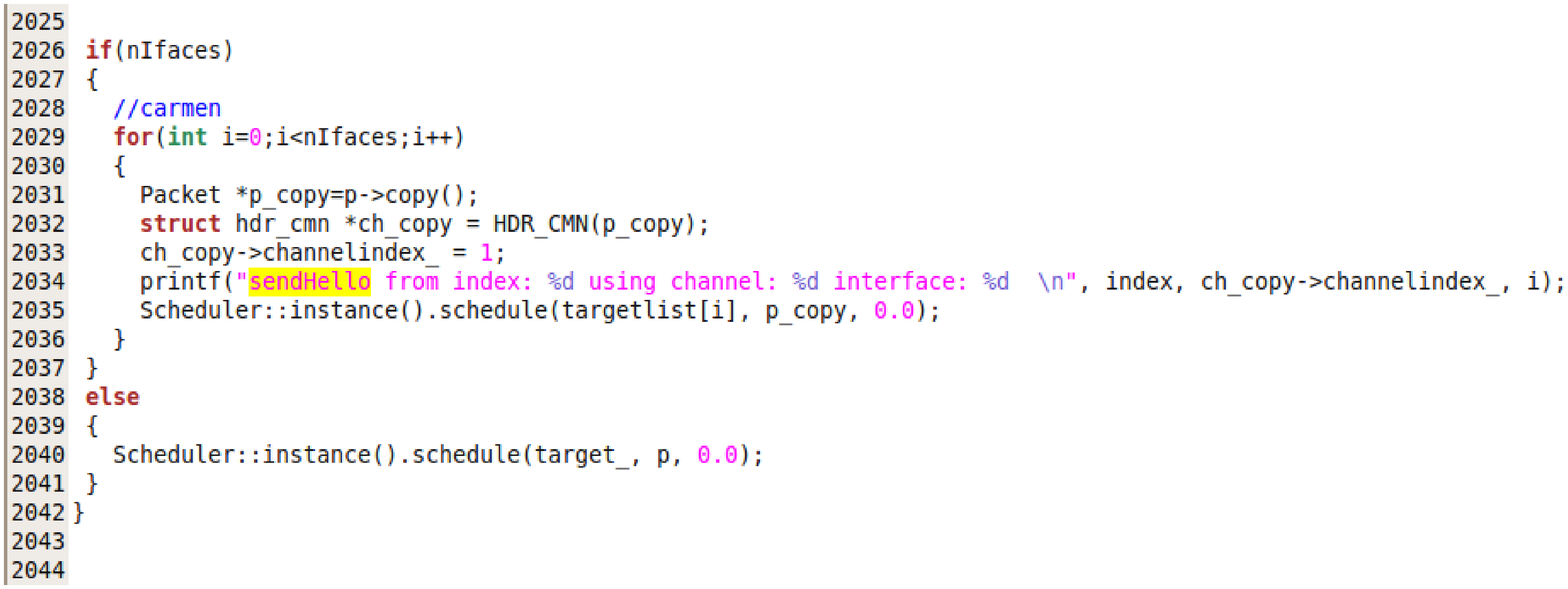}\vspace{-1.4cm}
    }
\caption{SendHello() function in the wcett.cc file.}
\label{fig_1}
\end{center}
\end{figure}

In order to send Hello packets on a specific channel over multiple interfaces, we need to assign the specific channel. This assignment of channel can be done by using the following variable:

$ch\_copy->channelindex\_$ 

This channel value will then used by the network interface (phy.cc file) to select a channel for sending the packets. We can check how packets are transmitting on a specific channel over multiple interfaces by adding some lines in SendHello() function of wcett.cc file (cf. Fig.~\ref{fig_1}). In Fig.~\ref{fig_1}, the lines from 2026 to 2041 send Hello packets over multiple interfaces. Note that if we do not specify the channel selection explicitly, as we mentioned on line 2033 (channel \# 1), nodes will send the packet automatically on the channel decided at the lower layers.

\begin{figure}[h]
    \begin{center}
    {
        \includegraphics[width=13cm, height=4.5cm]{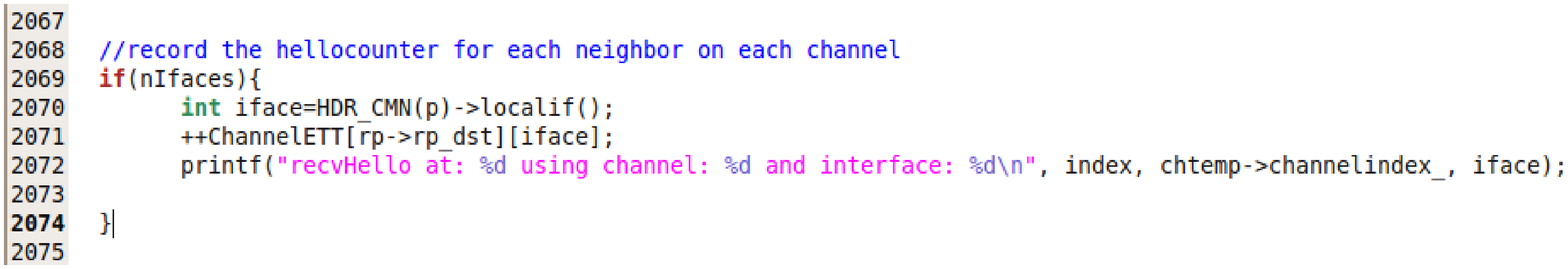}\vspace{-1.4cm}
    }
\caption{RecvHello() function in the wcett.cc file.}
\label{fig_2}
\end{center}
\end{figure}

We can also see how nodes receives the hello packets by using printf command in recvHello() function present in the wcett.cc file (cf. Fig.~\ref{fig_2}). \vspace{-0.5cm}

\subsection{Ways of Selecting the Channel}

%

There are two ways to select the channel value:

\begin{enumerate}
\item The routing layer decide the channel and communicate this information to lower layer. Note that this is done by adding a field in the packet header. Lower layer can use this channel decision value from the packet header. This channel selection decision value can be used to select the channel of the current node by making the following change in phy.cc file at the beginning of its recv() function:

          $nchannel = hdr->channelindex\_;$
          
which changes the current channel value to equal $channelindex\_$ value passed from WCETT routing layer or simply wcett.cc.

\item We can assign some channels to some specific nodes. For e.g. we can select the channel 0 for node 7's interfaces and for all other nodes, the selected channel value is 1. This could be done by adding the following lines at the start of  the recv() function in phy.cc 

        if $(node()->nodeid() == 7)$
        
        nchannel = 0;
           
        else
        
        nchannel = 1;

\end{enumerate}

If we make the above changes, channel 0 is assigned to node 7, while channel 1 is assigned to all the remaining nodes.


\subsection{Modifications in wireless-phy.cc file}

\begin{figure}[h]
    \begin{center}
    {
        \includegraphics[width=13cm, height=4.5cm]{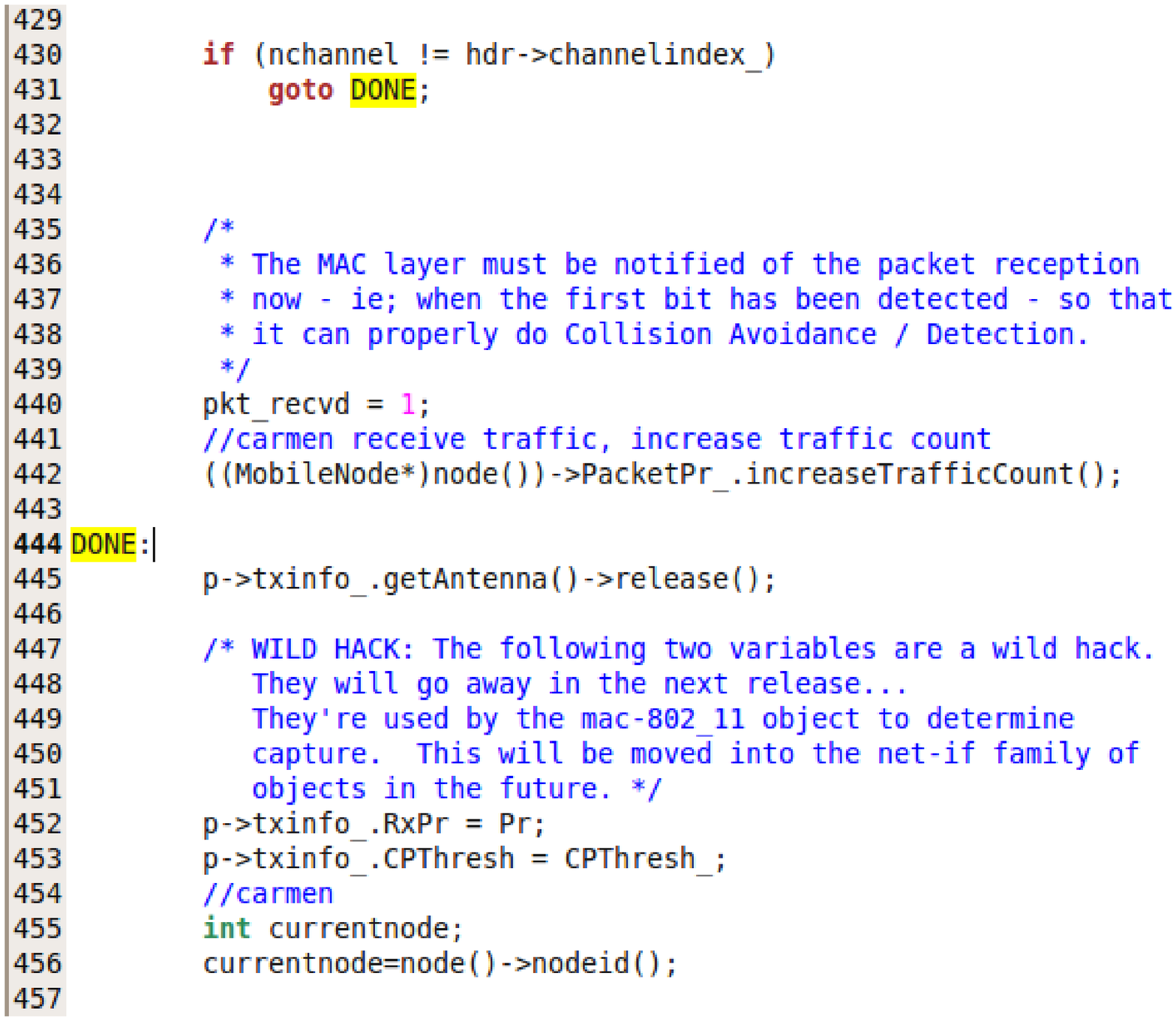}
    }
\caption{SendUp() function in the wireless-phy.cc file.}
\label{fig_3}
\end{center}
\end{figure}

In order to tune an interface based on the channel value passed from the routing layer (i.e. wcett.cc), we are required to make the following change in sendUp function of wireless-phy.cc file. The lines 430-431 (cf. Fig.~\ref{fig_3}) tunes the interface to the channel decision communicated from the routing layer.



%
%
%
%
%
%

\section{Conclusion}
In this tutorial, we showed that how broadcasting of packets can be done in multi-interface network setting in NS-2. Moreover, we also demonstrated through a working example that an interface can be tuned to a particular channel based on the channel value passed from the routing layer (that is wcett.cc). One can also perform complete broadcast by broadcasting over all the available channels. 

\bibliographystyle{IEEEtran}
\bibliography{aodvbib}

\end{document}